\documentclass[conference]{IEEEtran}
\usepackage{graphicx}
\usepackage{threeparttable}
\usepackage{ifpdf}
\usepackage{soul,color}
\usepackage{tabularx}

\usepackage{forest}
\usepackage{fancyhdr}

\ifpdf

\else
\fi
\usepackage{pgfplots}
\usepackage{array}
\usepackage{float}
\usepackage{graphicx}
\usepackage{subfigure}
\usepackage{url}
\usepackage{leftidx}
\usepackage{multirow}
\usepackage{array}
\usepackage{longtable}
\usepackage{rotating}
\usepackage{amsmath}
\usepackage{amssymb}
\usepackage{textcomp}
\usepackage{hyperref}
\usepackage{booktabs}
\usepackage{multirow}
\usepackage{ragged2e}

\usepackage{siunitx}
\usepackage{makeidx}  
\usepackage{cite}
\usepackage{graphicx}
\usepackage{blindtext}
\usepackage[export]{adjustbox}
\usepackage{verbatim}
\usepackage{enumitem}
\graphicspath{{figs/}}

\begin{document}
\bibliographystyle{ieeetr}
\title{A Novel Internet-of-Drones and Blockchain-based 
 System Architecture for Search and Rescue}


\author{
\IEEEauthorblockN{Tri Nguyen\textsuperscript{1}, Risto Katila\textsuperscript{2}, and Tuan Nguyen Gia\textsuperscript{2}}

\IEEEauthorblockA{
\textsuperscript{1}Center for Ubiquitous Computing, University of Oulu, Oulu, Finland\\
\textsuperscript{2}Department of Computing, University of Turku, Turku, Finland\\
Email: \textsuperscript{1}tri.nguyen@oulu.fi, and \textsuperscript{2}\{rijukat,tunggi\}@utu.fi\\
}}
\maketitle
\maketitle              

\begin{abstract}    

With the development in information and communications technology (ICT) and drones such as Internet-of-Things (IoT), edge computing, image processing, and autonomous drones, solutions supporting search and rescue (SAR) missions can be developed with more intelligent capabilities. In most of the drone and unmanned aerial vehicle (UAV) based systems supporting SAR missions, several drones deployed in different areas acquire images and videos that are sent to a ground control station (GCS) for processing and detecting a missing person. Although this offers many advantages, such as easy management and deployment, the approach still has many limitations. For example, when a connection between a drone and a GCS has some problems, the quality of service cannot be maintained. Many drone and UAV-based systems do not support flexibility, transparency, security, and traceability. In this paper, we propose a novel Internet-of-Drones (IoD) architecture using blockchain technology. We implement the proposed architecture with different drones, edge servers, and a Hyperledger blockchain network. The proof-of-concept design demonstrates that the proposed architecture can offer high-level services such as prolonging the operating time of a drone, improving the capability of detecting humans accurately, and a high level of transparency, traceability, and security.   

\end{abstract}


\IEEEpeerreviewmaketitle
\thispagestyle{fancy}
\pagenumbering{gobble}
\lhead{\scriptsize This is the accepted version of the work. The final version will be published in MASS virtual conference, 04-10 Oct 2021. ©2021 IEEE. Personal use of this material is permitted. Permission from IEEE must be obtained for all other uses, in any current or future media, including reprinting/republishing this material for advertising or promotional purposes, creating new collective works, for resale or redistribution to servers or lists, or reuse of any copyrighted component of this work in other works.}

\section{Introduction}\label{introduction} 
Search and Rescue (SAR) is a complex operation that combines different activities and actions from multidisciplinary professionals~\cite{coppola2006introduction}. The main target of SAR is to search for missing persons and provide aid for persons who are in danger in different situations such as earthquakes, floods, storms, and snowstorms~\cite{coppola2006introduction}. SAR can include different missions such as mountain, cave, and maritime SAR. Each SAR mission can require a different number of specific staff and equipment~\cite{coppola2006introduction}. To perform a maritime SAR mission, staff with a rescue boat and a rescue helicopter will go to the location where the last signal or evidence is found. It is challenging to complete SAR missions in many situations due to many reasons such as large search areas, harsh environments (i.e., wind, heat, or GPS unsupported), and strict time requirements~\cite{coppola2006introduction}. With the development of technologies such as wireless communication and cameras, SAR missions become less complicated as the technologies can help minimize the search time and improve the performance of a rescue team. For example, it is difficult for rescue teams to search for shipwreck survivors, especially at nighttime. However, high-resolution RGB and thermal cameras and advanced image processing approaches can help detect the survivors more easily~\cite{agrafiotis2016real}. Although the existing SAR systems using the technologies above have provided many advantages for SAR missions, they still have several limitations. For instance, it is challenging to reduce the searching time or cover the large searching areas in a short period. There is a need for more enhanced SAR systems that can both overcome the challenges of the existing SAR systems and help improve the possibility of finding victims of accidents and disasters. 

Drones and UAVs have been widely used in many applications such as surveillance, geographic mapping, traffic control, and power line inspection because they offer many advantages such as cost-saving, large coverage areas, and ease of deployment~\cite{zaheer2016aerial,abushahma2019comparative}. For example, drones and UAVs can be equipped with different sensors and cameras to cover a large area and different terrains efficiently. Many drones and UAV systems have recently been developed to serve SAR missions, including cave, avalanche, and maritime SAR~\cite{qingqing2020towards,queralta2020autosos}. However, drones and UAVs still have some limitations, such as limited computation resources and battery capacity. Therefore, they cannot operate for a long period. These limitations affect the performance of the rescue team. One of the suitable approaches for extending a drone's operating time is to apply computation offloading that can switch tasks from a drone into a ground control station. However, when the connection between the drone and the ground control station is disconnected, the offloading approach cannot be carried out.    

Internet-of-Things (IoT), consisting of different technologies such as cloud computing, sensing, and communication, offers many benefits, including real-time and remote monitoring~\cite{tcarenko2016energy,ali2019intelligent}. However, IoT still has some limitations related to latency-sensitive issues and data transmission~\cite{rahmani2017exploiting}. These limitations can be overcome by edge computing that can be described as computing at the edge of the network closer to the location where the data is collected. Edge computing helps bring the cloud computing paradigm to the edge of the network and supports different fundamental features that cloud computing cannot provide. The combination of IoT and edge computing can help enhance the energy efficiency of hardware-constrained devices (i.e., smart sensors or drones) and provide advanced services such as real-time micro-service.  

Blockchain technology is a concept of building a unique database of a decentralized system. In detail, a decentralized system consists of many participants working together to form a unique state instead of based on a single participant deciding the system state~\cite{ali2016blockstack}. Therefore, blockchain technology demands to replicate any system changes to all participants and form agreements with operations of the system. Blockchain offers many advantages such as enhanced security, transparency, traceability, efficiency, and automation~\cite{ibm}. For example, blockchain uses cryptographic hash functions to form a chain of data blocks linking by storing the ancestor's hash value. By participating in a blockchain network, all the participants with permission can access the same information simultaneously. Due to blockchain's merit, blockchain is applied in many applications such as healthcare, cryptocurrency exchange, and trading~\cite{gia2019artificial,nakamoto2008bitcoin,nawaz2020edge,allouch2021utm}.

SAR missions need to be transparent, efficient, traceable, secure, and precise. Particularly, to achieve a high level of effective collaboration and efficiency, each participant with permission, such as remote medical staff at the hospital, can access the same information related to a SAR mission (e.g., the status of a victim) simultaneously as the rescue team. This helps improve possibilities to save the victim. For example, all the necessary equipment and medical staff for an operation can be carefully prepared. When a victim is brought to the hospital, the operation can be carried out immediately. In addition, it is required to protect sensitive data related to a victim and the SAR mission from persons who are not related to the mission. Due to the blockchain's benefits,  blockchain can be one of the suitable approaches for fulfilling SAR requirements in terms of security, transparency, and traceability.

It is challenging to develop drone and UAV-based systems for SAR missions as they often have many strict requirements such as low latency, high precision, large coverage area, and security. Therefore, this paper proposes an advanced Internet-of-Drone (IoD) system architecture integrating edge computing and blockchain for SAR missions. The proposed architecture consists of different types of drones, including small drones and big drones that have different sizes, computation resources, flying capabilities, sensors, cameras, and prices. The proposed architecture provides a solution that helps overcome the challenges of the drone and UAV-based system for SAR and offer rooms for further development such as scalability and fault tolerance. Remarkably, many existing UAV-based systems need a long period to completely cover the large searching areas and find many people simultaneously in a SAR mission. In addition, it is costly to build an advanced UAV-based system that can search a large area with minimum latency. In this paper, a maritime SAR application is focused. Therefore, the proposed system architecture will be implemented for targeting maritime SAR missions.



The structure of the paper is as follows: Section~\ref{sec:rela} provides related work. Section~\ref{sec:blockchain} mentions a background of blockchain technology. Section~\ref{sec:architecture} shows the proposed system architecture, while Section~\ref{sec:offload} discusses the preliminary knowledge of computation offloading. Section~\ref{sec:setup} presents the system setup and implementation, and the result of experiments is in Section~\ref{sec:experiment}. Finally, Section~\ref{sec:conclu} indicates future work and conclusion.  
\section{Related work} 
\label{sec:rela}

Recently, drone and UAV-based systems for SAR missions have been developed due to the benefits of UAVs and drones. However, the number of these approaches is limited. Most of the works target a single drone or a drone group that directly connects to a ground station center. Drone and UAV-based approaches for SAR missions have not simultaneously considered different aspects of drones/UAVs, remote monitoring, computation offloading, and blockchain. In general applications, most of the works consider Internet-of-Drones with either offloading or blockchain. Therefore, the paper discusses drone and UAV-based architectures for SAR missions, drone architectures for computation offloading, and drone architectures utilizing blockchain. 


The authors~\cite{Scherer2015} presented a modular architecture for the UAV system for SAR. The system architecture consists of UAVs, viewer base stations, and control base stations. A viewer base station and a control base station have a user interface that allows access to sensed data and controls different aspects of the system, respectively. The results show that the proposed architecture can be applied for different scenarios with different levels of autonomy. A drone of UAVs can join or leave a system network without compromising the quality.

Koubaa et al.~\cite{koubaa2020deepbrain} introduced an IoD architecture for computation offloading. The architecture with drones, edge, and cloud allows for offloading tasks from drones over the Internet to cloud servers powerful in computation and memory storage. Mainly, images captured by a drone can be firstly processed at edge servers and then at cloud servers. The result demonstrates that offloading can help reduce average latency.  

In \cite{wang2020agent}, the authors proposed an agent-based architecture for computation offloading. The architecture has three layers, including user devices, access network, and core network. A user device layer consists of sensor devices collecting and forwarding data to the access network that comprise UAVs and edge servers for local data processing or data transmission to cloud servers. The results show that architecture can be applied for various applications in different fields.

In \cite{jung2017acods}, the authors presented a computation offloading system architecture to extend the flight time of surveillance drones. The architecture includes a response time prediction module, task offloading module, and remote task execution module. The response time prediction module at drones is responsible for collecting data such as wireless communication and mobility factors, while the task offloading module is for making an offloading decision based on the collected data. If computation offloading is made, the task will be sent to the remote task execution module that is carried out at the ground control system. The results demonstrate that the proposed architecture help increase the flying time of the drone.  

Hou et al. \cite{hou2020distributed} described a swarm of drone architecture utilizing fog computing for task offloading. The architecture supports cloud-based computation offloading and task allocation optimization. In the architecture, each drone with a fog node has connections with its adjacent drones or UAVs. After local processing of the particular tasks at the drones is made, the node and its adjacent drones are allocated the tasks via the task allocation link. A task allocation optimization is built to reduce a swarm of drones' energy consumption while satisfying the latency and reliability requirements. The task allocation is based on the fast proximal Jacobi alternating direction method of multiplier proposed in~\cite{deng2017parallel}. First, a task allocation optimization problem is split into smaller sub-tasks in which each drone is responsible for the specific sub-tasks depending on their local status information. When comparing with other task allocation optimization algorithms, including the centralized convex optimization algorithm and the heuristic algorithm developed in~\cite{hou2019fog}, the simulated result proves that the proposed algorithm is more efficient in optimization capability, expandability, and convergence rate. Compared with other state-of-the-art drone and UAV-based computation offloading approaches, the proposed fog approach has been more comprehensively evaluated in terms of reliability performance.

In \cite{allouch2021utm}, the authors proposed an architecture using a blockchain network for IoD. The architecture utilizes a lightweight blockchain network to offer secure communications between participants of the system. The result shows that the proposed architecture can help improve data sharing between UAVs and their control systems.

The authors~\cite{sharma2017socializing} presented an architecture utilizing drones and blockchain to improve inter-service operability in terms of trust. The authors show the benefits of the proposed blockchain architecture towards the traditional centralized drone-based systems, although the architecture is not actually implemented. 




\section {Preliminaries: Blockchain technology} \label{sec:blockchain}


\begin{figure}[t!]
	\centering
	\includegraphics[scale=1]{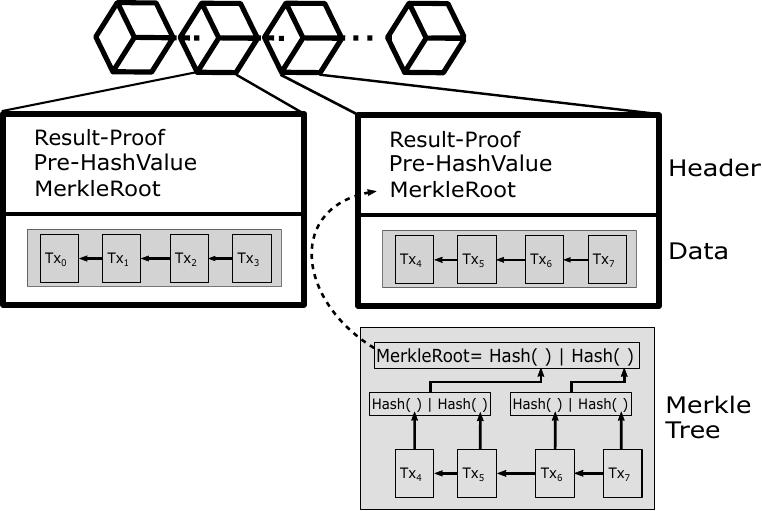}
	\caption{The blockchain structure}
	\label{fig:blockchainStructure}
	\vspace{- 10pt}
\end{figure}

The blockchain succeeds in forming a decentralization by gaining the consistency of the database replication. In detail, the consistency ensures the correctness of the system with a unique state from different replications; however, obtaining consistency in a decentralized environment is difficult because of a demand of reliable communication among participants and agreements with state updates. Therefore, the consistent feature of a decentralized organization requires a specific structure and a consensus mechanism working along with cryptographic schemes.
    
Blockchain's structure observes as a chain of data blocks extending by time and linking by the hash pointer concept. The blockchain structure consists of two main parts: blocks and transactions, as Fig.~\ref{fig:blockchainStructure}. The transaction indicates a primary type of data in the blockchain. Also, a block wraps a set of chronological transactions in its data part. Besides, each block forms a header containing metadata, including a hash value of its predecessor, a root of Merkle Tree data organization~\cite{merkle1987digital}, and a puzzle solution. As a result, if there is any modification from the block predecessors, the current block can detect from the hash pointer. Even if there are changes in the data part, the current block recognizes the storage of the Merkle root tree.
    
Consensus mechanisms play a crucial role in blockchain-based system formation. In particular, consensus provides a solution of communication to gain the agreement between network participants. The consensus for a blockchain system is analyzed by Xiao et al.~\cite{xiao2020survey} into five components, including block proposal, information propagation, block validation, block finalization, and incentive mechanism. In detail, the block proposal is to generate a data block candidate that is then broadcast to other participants in the information propagation. After receiving the block candidates, the participants verify the block candidate at the block validation and then find a suitable branch to attach the validated block at block finalization. Finally, the incentive strategy is to reward participants contributing to the system or even punish participants who do not obey the protocol. For example, Bitcoin~\cite{nakamoto2008bitcoin}, the first blockchain-based cryptocurrency, proposed Nakamoto consensus based on the Proof-of-Work concept. A Nakamoto consensus's participant forms a block candidate after solving a computation puzzle. The block candidate is then shared with others for verification of the correctness before attaching the current blockchain. Once the consensus round is reached, the participant forming the accepted block is rewarded. In this way, the system encourages the participants to join in the consensus process.
    
An exciting application of blockchain is the blockchain-based smart contract. After the proposal of blockchain technology with cryptocurrency use cases, Vitalik established Ethereum~\cite{wood2014ethereum} as a blockchain-based smart contract to utilize Ethereum Virtual Machine (EVM), a Turing-complete machine for arbitrary system activities. Therefore, except for storing and forming a blockchain structure with transactions, blockchain-based smart contract, as an example with Ethereum, has application binary interface two-direct converting between Solidity (a programming language in Ethereum) function call and EVM byte code. Therefore, each Ethereum participant organizes an EVM that executes defined smart contracts and receives requests as function calls. As a result, the blockchain-based smart contract is observed as the next evolution of blockchain technology use case for a decentralized autonomous organization (DAO). With this concept, smart contracts can work autonomously, and those operations are stored with a blockchain among participants. Although Ethereum is the first blockchain-based smart contract for permissionless blockchain, our design is based on the use of permissioned blockchain to ensure the authentication of participants in the network.
    
With blockchain concepts, many blockchain-based applications have been gaining the advantage of the decentralized environment. One of the most exciting benefits of blockchain technology is decentralization, supporting single-failure tolerance. Also, blockchain technology provides participants with trustworthiness through an immutability and transparency feature by which the system can form audits. Another attraction for the proposal is the utilization of smart contracts to form DAO from which the system can swiftly react and operate issues and even self-organization. For instance, when one small drone or big drone's battery is going to end soon, the system, after gathering the degree of the specific drone, can find another drone to replace that one. In another scenario, if one of the edge servers was crash or even in an attack, the system still operates typically with other edge servers.
    

\section{Architecture}  \label{sec:architecture}

\begin{figure*}[t!]
	\centering
	\includegraphics[width=0.75\textwidth]{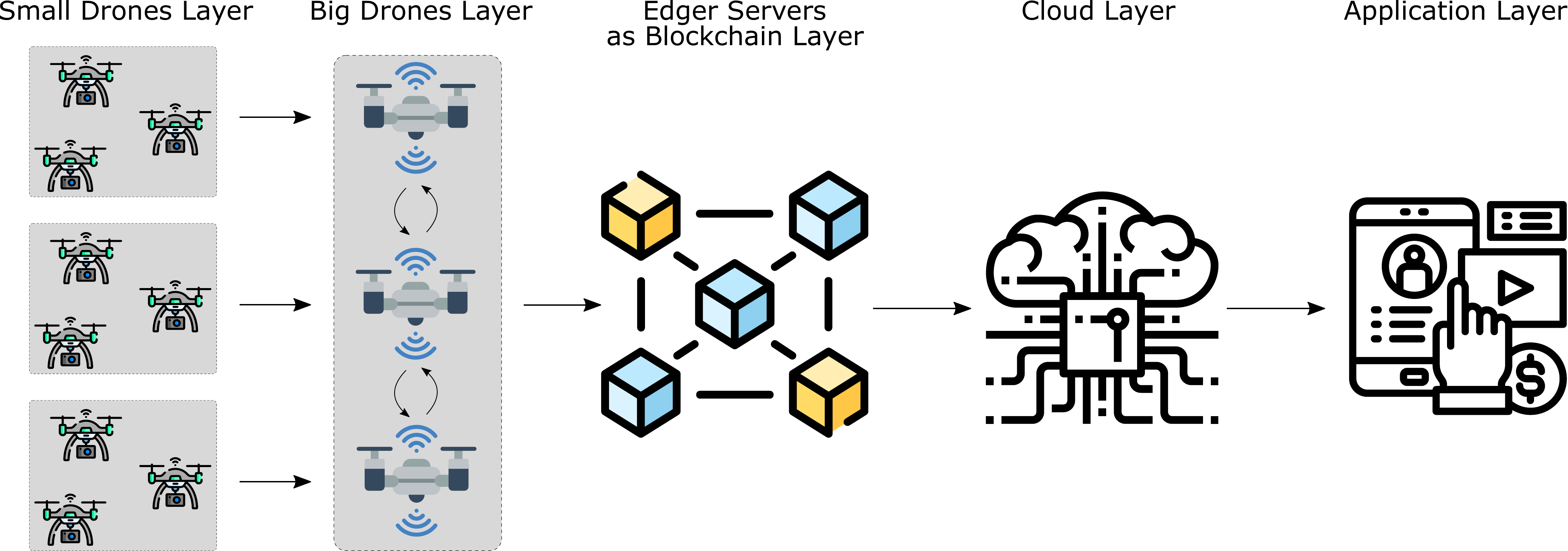}
	\caption{The proposed IoD architecture with integration of edge computing and blockchain}
	\label{fig:generalFig}
\end{figure*}

\begin{figure}[t!]
	\centering
	\includegraphics[scale=0.25]{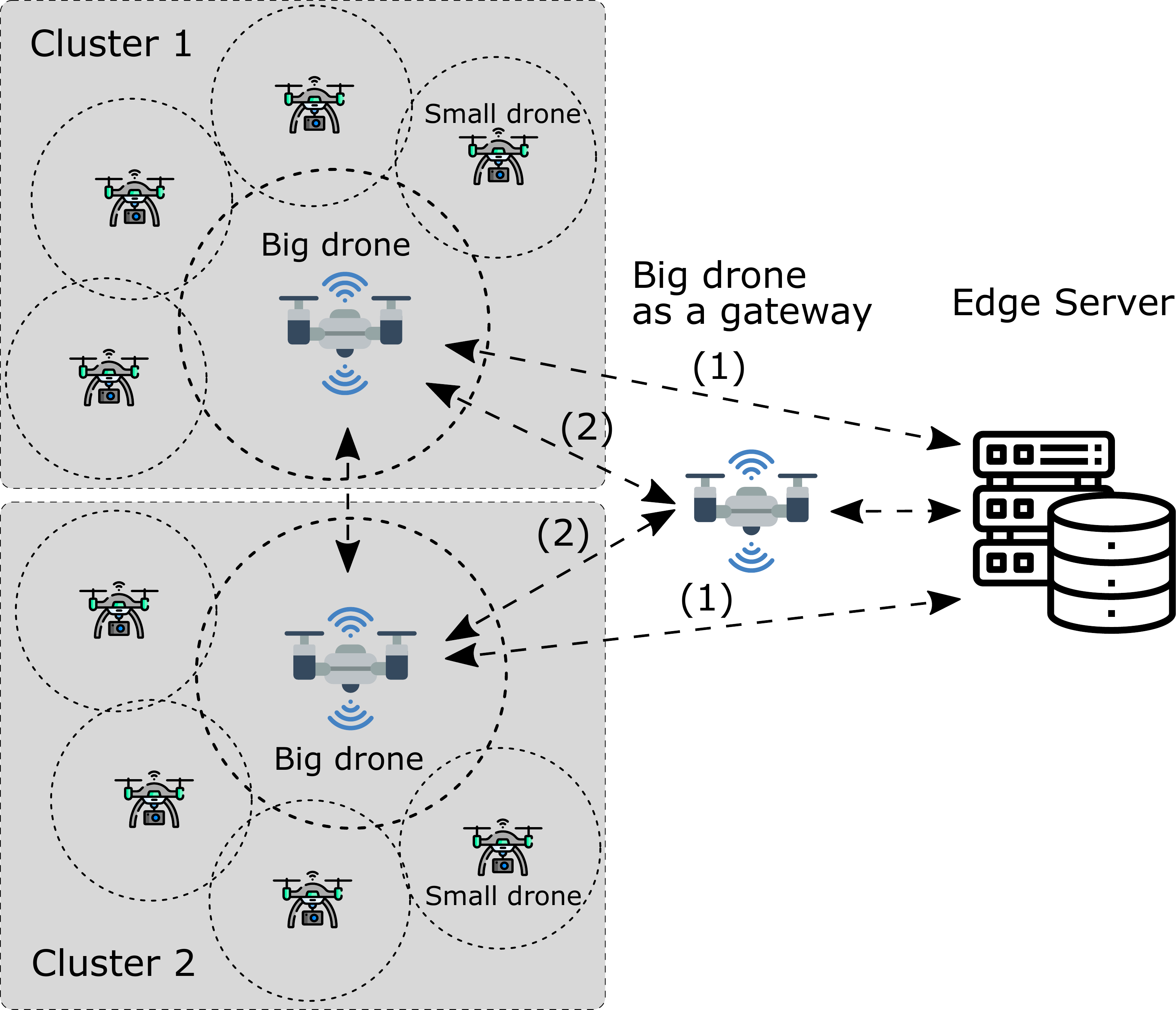}
	\caption{The cluster of drones and the communication of the drones in the proposed architecture}
	\label{fig:detailgeneralFig}
	\vspace{-10 pt}
\end{figure}

The proposal architecture shown in Fig.~\ref{fig:generalFig} consists of five layers, including small drone layer, big drone layer, edge servers as blockchain layer, cloud layer, and application layer. The small drone layer can consist of heterogeneous small drones, small, fast, resource-constrained, and low-cost compared with a big drone. Due to small-load and low battery capacity limitations, the small drone is often equipped with a small set of cameras and sensors. Depending on the application, the cameras and sensors need to be carefully chosen. For example, a high-speed RGB camera (e.g., Intel Realsense D435) and sensors (e.g., digital barometric pressure sensor for air pressure) can be used to track object applications. In the proposed architecture, the small drones work as a cluster in which the small drones work independently and cover a particular search area depending on the drone's operating time. The cluster of small drones and the communication of the drones are shown in Fig.~\ref{fig:detailgeneralFig}. More small drones should be added to a cluster to cover the larger area, or the drones of the cluster should fly higher. However, increasing the flying height can cause higher energy consumption and reduce the accuracy of human detection. The number of small drones can vary depending on the applications or missions. Each cluster has a leader, which is a big drone. Small drones do not communicate with each other, but they only communicate directly with a cluster leader via Wi-Fi. The small drones are responsible for capturing images or videos which can be local pre-processed or kept intact before being sent to the leader, depending on the offloading decisions. 


The system architecture can have several clusters of drones. Cluster leaders communicate with each other via Wi-Fi and form a big drone layer. The big drone is equipped with powerful computation resources capable of processing a large amount of data quickly in a parallel manner. Embedded boards with GPU having multiple cores (e.g., NVIDIA Jetson Xavier NX with 384 cores) can be applied. Depending on the applications and the budget, the big drone can be equipped with or without cameras.
In most cases, a big drone can consist of different types of cameras such as RGB and thermal cameras that are very useful in searching people at night. In general, the price of the big drone without the camera can be from 2k euros to 8k euros, while the big drone with RGB and thermal cameras can be from 10k to 15k euros. The price of the big drone with cameras is higher than the small drone's price (e.g., 2k euros). To improve the accuracy of human detection, it is recommended that the big drone be equipped with high-speed cameras, including RGB and thermal, that can capture high-resolution images and videos at night. Depending on the drone type and price, the big drone can operate with different flying times (e.g., DIJ s1000 with 15-minute flying time and DJI Matrice 300 with 55-minute flying time without any load). Wi-Fi is used as the primary communication protocol between the big drones and an edge server placed at a rescue boat in the proposed architecture since Wi-Fi can support high bandwidth and high data rate transmission. 

The proposed system architecture supports offloading between cluster leaders. For example, suppose one cluster has much fewer members than another cluster. In that case, a cluster leader with fewer members may have some more available resources when these leaders' specification is the same. Correspondingly, a cluster leader with less available resources can offload some of its tasks to its adjacent leaders having more available resources. This offloading feature only works if the connection between these leaders is good and adjacent leaders have available resources. However, this offloading feature is not enabled in most situations as it will cause some challenges (e.g., complex management and extra communication overhead). Although the offloading feature is off, cluster leaders can communicate with each other to exchange critical information and deal with a bad connection between a cluster leader and an edge server. Notably, a cluster leader can provide multicast to all adjacent cluster leaders to inform the situation or share information (e.g., human detection results and the current location). Then, the adjacent cluster leaders will forward the information to an edge server placed on the rescue boat and send back the edge server's temporary location to the cluster leader, who cannot connect with the edge server temporarily. This will help the leader to know its current position and distance toward the edge server. Correspondingly, a rescue boat may relocate into another position, or the leader may adjust the transmission power to maintain the stable connection. This can ensure that the system can work adequately; even GPS has temporary issues in specific areas.

The proposed architecture can support different types of human detection algorithms. However, it is required that the algorithm for detecting humans at a big drone/a cluster leader must satisfy some levels of accuracy and low latency requirements defined by system administrators. For example, the 90\% accuracy level and maximum latency of 500 ms can be pre-defined. After the human detection algorithm detects a person at a cluster leader, the leader can verify the case by using its camera to capture the images or video of the areas where the person is detected. In this situation, a leader's camera will be zoomed to capture the images and videos. Then, the acquired images and videos are processed to detect the person. After that, the cluster leader can fly closer to the area where a victim is detected to capture images and video sent to an edge server for further detection by advanced algorithms and rescue staff. This improves the possibility of finding the victims. If a cluster leader is not equipped with any camera, another approach can be applied. For instance, after the human detection algorithm at the leader detects a victim from images or videos sent by a small drone, the leader can send commands to order the small drone to fly closer to the victim. Correspondingly, better images and videos related to the victim can be captured. The acquired data is then forwarded to an edge server to further detect by complex algorithms or rescue staff. 


In the proposed architecture, if a human detection algorithm running in a big drone detects a person, then the images and the video will also be forwarded to an edge server. Otherwise, if a person is not detected, big drones will not send images and videos to the edge server in real-time because of several following reasons. Firstly, the good connection between big drones and an edge server cannot be easily maintained all the time because an edge server is on a non-stationary rescue boat. In many cases, the boat may go out of the connection range of some big drones. For another reason, this can help save energy consumption of the big drone. Particularly, if each small drone sends a video of 30 frames per second (fps) to a big drone and a cluster has six small drone members, the big drone needs to send 180 fps to an edge server. The big drone can process the received data and send the pre-processed data, which can be much smaller, but this still causes latency and consumes energy. When a Wi-Fi dongle is turned on, it can consume around 400-500 mJ per second. Although the big drones do not send images and videos in real-time to an edge server, the data stored locally in the big drones are transferred to the edge server when the big drones come back to the boat for charging the battery or replacing a new battery. This helps build a dataset for training deep learning-based models such as YOLOv4 \cite{bochkovskiy2020yolov4}. In addition, at the edge server, the advanced algorithms and staff from a rescue team can double-check the received images and videos to avoid missing any victim.          

The proposed architecture also supports the case of a big drone as a gateway. In case of a harsh environment such as noises or strong wind, the system administrators or the rescue team can decide to use an extra big drone as gateways which are responsible for receiving data, including images and videos from cluster leaders and forwarding the data to an edge server. This approach helps avoid missing data. In another situation, the gateway can be used as a backup plan for maintaining the wireless communication or an extra aid device for helping the victim. In case of where many victims have been detected in different areas the aid device can provide some necessary equipment or aid packages to the victim while waiting for the rescue boat rescuing other victims.

An edge server is placed at a rescue boat in this architecture. The edge server connects with big drones via Wi-Fi while connecting with the Cloud server via 4/5G. The edge server can work without considering the impact of energy consumption on the quality of service as it can be supplied from the rescue boat. 
The edge server has substantially more computational resources and data storage compared to the small and big drones.  Mainly, an edge server can store a large amount of data (e.g., a few Terabytes of data). The edge server is powerful enough to train some deep learning models for human detection. 

The cloud layer consists of cloud servers and cloud services. For example, cloud servers and services provided by Google, Amazon, or Microsoft can be used. The cloud servers can store a large amount of data for an extended period. When more data is stored, additional storage can be added easily without any effort. The server can perform heavy computations, and big data analysis, such as training deep learning models for artificial intelligence (AI) approaches to remove noises and detect humans accurately.
Similarly, the resources such as CPU (central processing unit), GPU (graphics processing unit), and memory can be added easily when necessary. Cloud servers also offer push notifications for informing a responsible person or a group of persons in real-time by sending text messages or notifications in a mobile app. A user with permission can use a mobile app connected to cloud servers to acquire information related to the search and rescue missions.  


The big drones then consider authentication for small drones in the proposed architecture before submitting to the servers. The servers can be edger servers that are distributed in a wide area. Those servers synchronize together to form a unique blockchain that stores information. The blockchain information is an audit for later study such as audit or analyzes the danger area. For example, from victims' data in dangers and current tasks of rescue teams such as (Global Positioning System) GPS and level of danger, the blockchain system provides optimized tactics to gain efficiency. As a result,  with information from rescue teams nearby, the blockchain system supports rescuers who quickly react to dangerous accidents.


The architecture satisfies requirements from the less powerful devices, and the blockchain has to be handled by powerful servers. Remarkably, the formation and management of a blockchain require at least a number of communication between the participants and even a massive computation for consensus process, cryptographic schemes, and verification. For example, if the blockchain formation is based on Proof-of-Work consensus as Bitcoin, the robust computation is an essential parameter for blockchain management.  Therefore, powerful servers, especially edge servers, have to implement the blockchain layer in our consideration. 

The blockchain layer controls the blockchain by agreeing and storing information from the system. In detail, the layer is a network of edge servers communicating among them to form a unique chain of data blocks to gain the same system state. Each edge server collects a bunch of data delivered by big drones before packing up in a block candidate for the blockchain extension. If the block candidate satisfies the requirement from the correctness of the system, this block gains agreement from other participants for attaching the current blockchain. Despite the widespread use of Proof-of-X consensus~\cite{tschorsch2016bitcoin} in blockchain-based systems, our blockchain-based architecture desires the use of Raft~\cite{ongaro2014search}. Raft is a leader-based consensus that means to select the next proposer for a block candidate; the system requires an election algorithm instead of a powerful computation to solve a puzzle as Bitcoin Proof-of-Work~\cite{nakamoto2008bitcoin}. Therefore, Raft is an efficient consensus for the proposed architecture.
Since blockchain is a distributed database, the proposed blockchain system maintains system operations and information. The blockchain contains participants' information, a smart contract system, and transaction execution. The blockchain stores drone information, including multimedia data, location, power, and metadata. Hence, the system can easily and quickly react to any issues through this information, such as the low battery of drones, crashes of drones, lousy weather, and obfuscation of specific areas. 

    \section{Computation offloading}    \label{sec:offload}
    
    In a UAV-based application, computation offloading can be described as switching computation (e.g., data processing, image processing for human detection) from a drone to other places, such as an edge server or a cloud server to save energy consumption drone or reducing the system latency. The proposed architecture supports several types of computation offloading, including task offloading from small drones to a big drone, offloading from big drones to an edge server placed at a rescue boat, and offloading between the big drones. However, the cases of computation offloading from a big drone to an edge server and computation offloading between the big drones are not the main offloading type because they hardly occur or do not provide many benefits. Therefore, in our paper, these offloading types are often disabled. Depending on the specific applications, these offloading types may be more useful.
    In the main case, computation offloading from small drones to big drones is carried out to extend the flight time of small drones, minimize the system latency, or maintain the accuracy level of human detection. Offloading consists of policies deciding that a task is processed locally at a small drone or offloaded for remote processing at a big drone. The general policies are rules based on results that compare the total energy consumption of performing a task and sending the result to a big drone with energy consumption of task offloading to a big drone. In general, due to the limited resources of a small drone, such as low battery capacity and low-speed CPU, it takes a long period to perform data processing for detecting a human from images or videos. Accordingly, most of the image processing tasks are offloaded into big drones. There is a trade-off between the latency and accuracy of human detection at big drones. For example, a big drone can use advanced and complex human detection algorithms (e.g., deep learning-based or machine learning-based algorithms) to achieve human detection accuracy. However, it may take a significant latency to run these algorithms. In this paper, several widely used image processing algorithms for human detection, such as Haar Cascades~\cite{viola2001}, Histograms of Oriented Gradients (HOG) based approach~\cite{dalal2005histogram}, YOLOv3-tiny~\cite{adarsh2020yolo}, YOLOv4~\cite{bochkovskiy2020yolov4}, and YOLOv4-tiny~\cite{jiang2020real} have been applied to demonstrate the policies. In addition to general policies, the system can apply specific policies for particular cases or scenarios. For instance, when a connection between a small drone and a big drone is interrupted or disconnected, the small drone can run the task locally or store the images locally based on specific policies which are more prioritized than the general policies.

\section{Setup and Implementation}  \label{sec:setup}
\begin{figure*}[t]
	\centering
	\begin{minipage}{1\textwidth}
	\centering
            \includegraphics[width=.2215\textwidth]{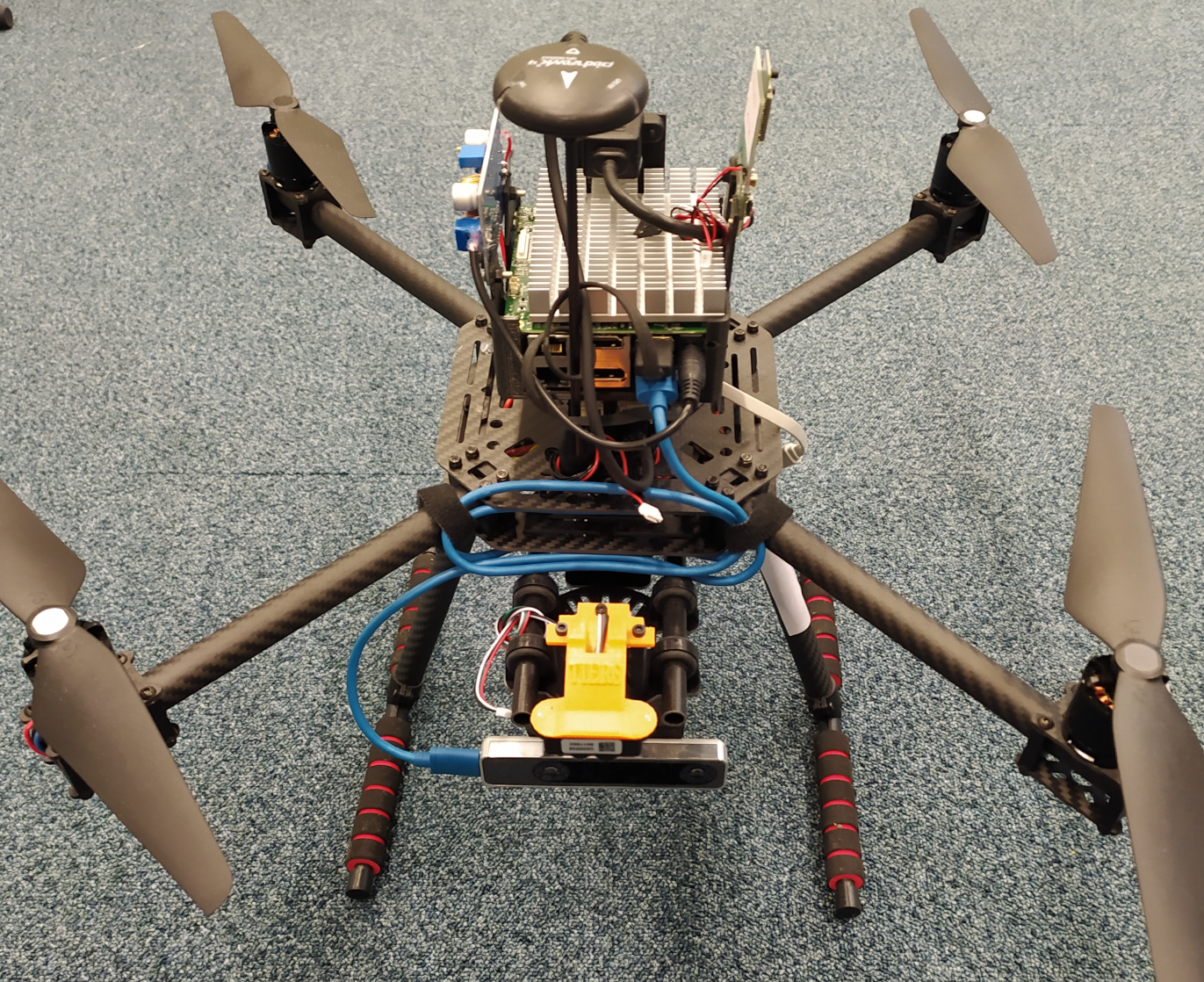} \hspace{102pt}
        	\includegraphics[width=.2656\textwidth]{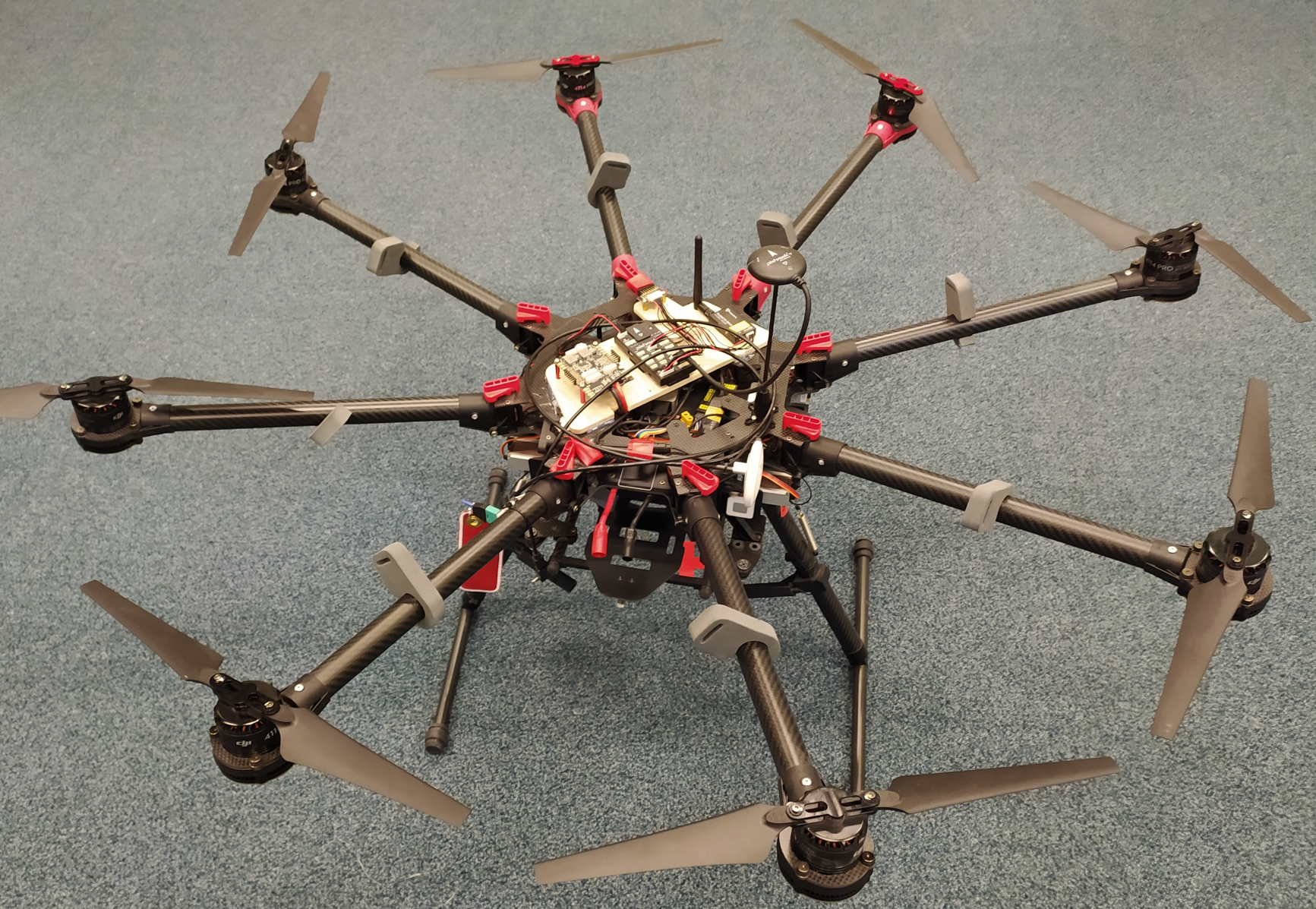}

        	\begin{FlushLeft}
        \hspace{102pt} (a) Small drone \hspace{124pt}
        	(b) Big drone (Customized S1000) \hspace{45pt}
        	\end{FlushLeft}
        	\includegraphics[width=.369\textwidth]{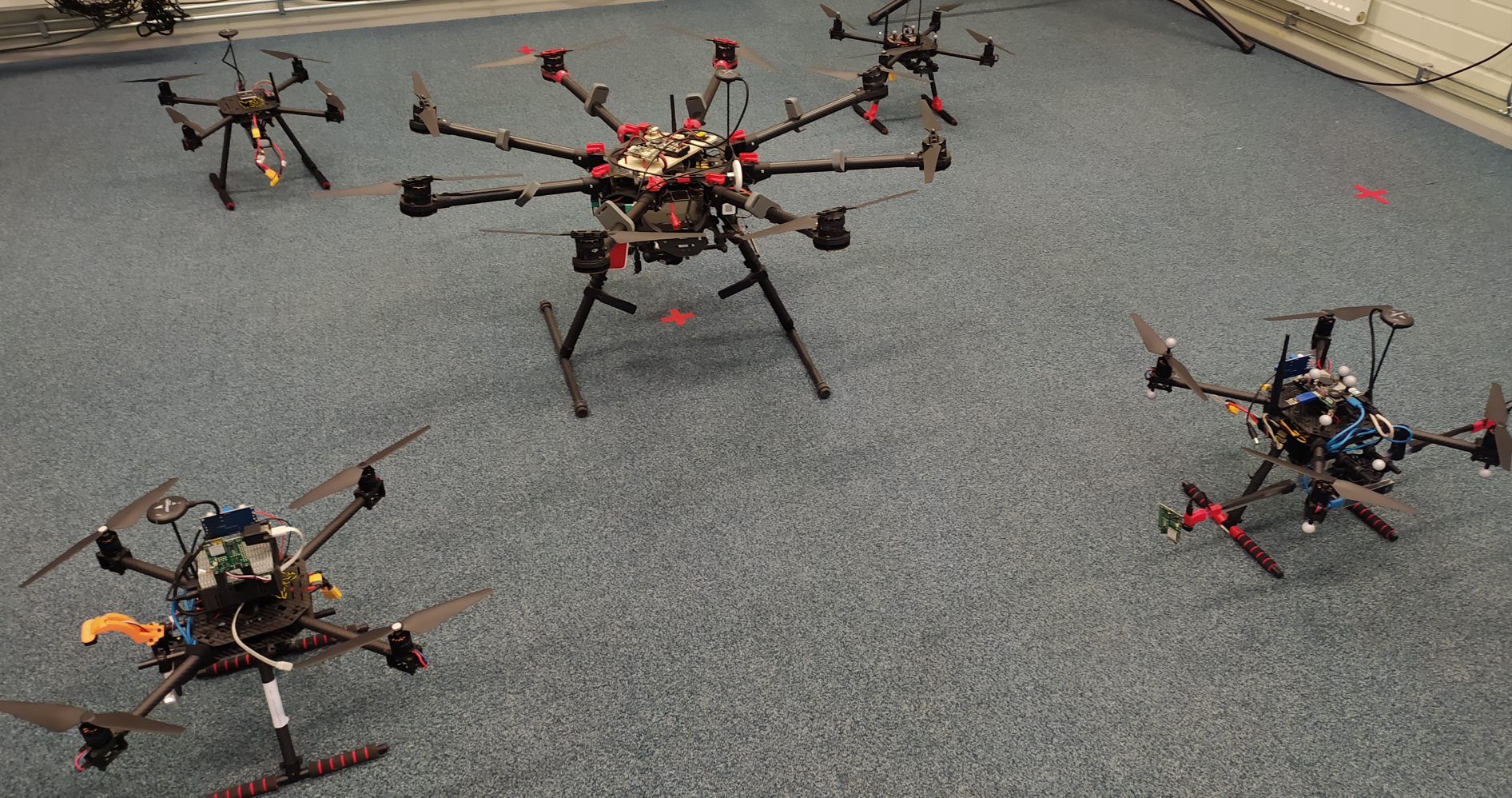} \hspace{45pt}
        	\includegraphics[width=.348\textwidth]{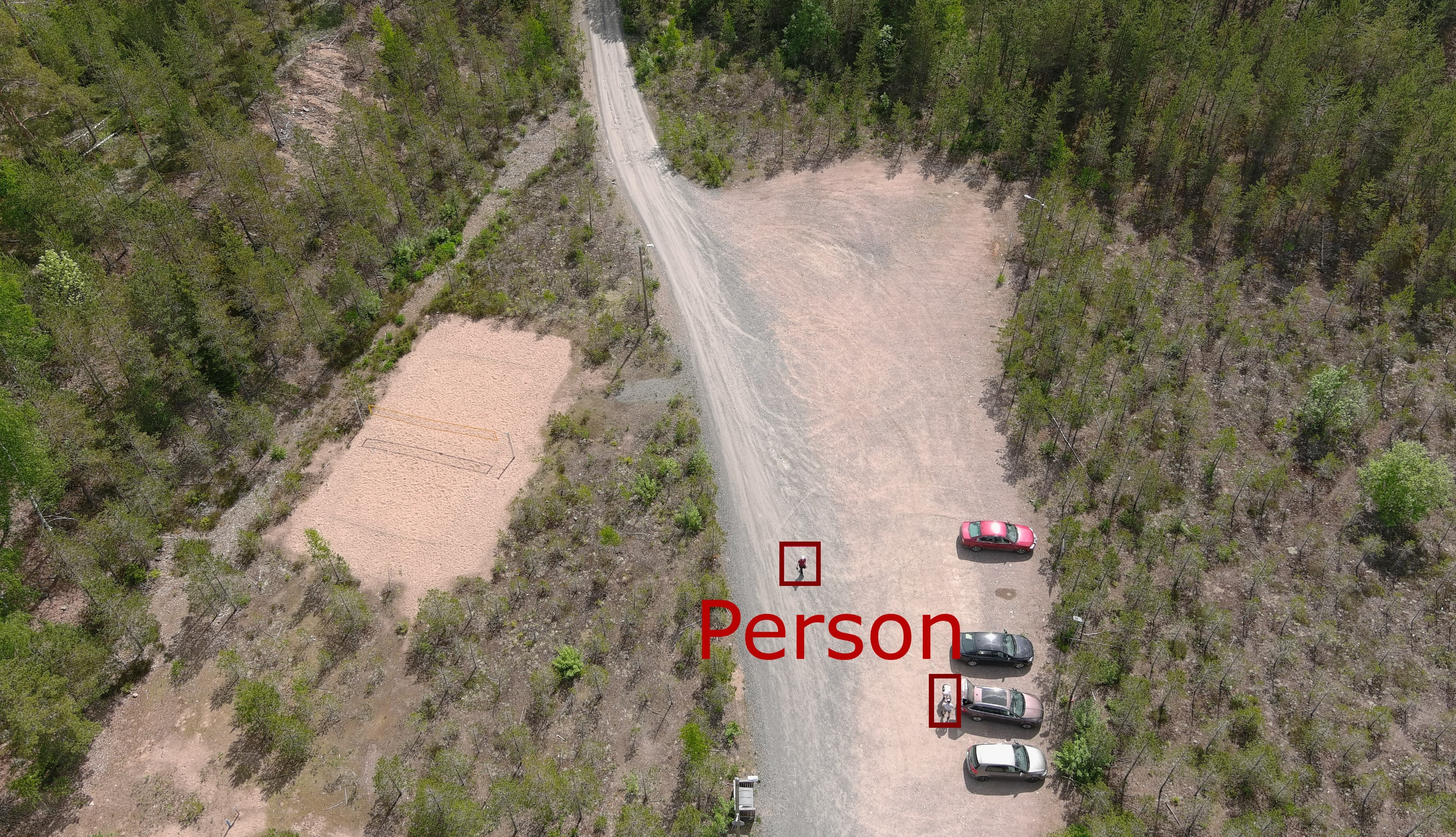}\\ 
        		\begin{FlushLeft}
        \hspace{80pt} \hspace{21pt}
        	(c) Cluster of drones \hspace{68pt}(d) An image captured by a drone (Autel robotics Evo II)\\ \hspace{265pt}with a 6K resolution camera at 60m height 
        
        	\end{FlushLeft}
    \end{minipage}
\caption{Examples of small and big drones, and a cluster of drones}
	\label{fig:drones}
\end{figure*}

The proposed architecture is implemented entirely from small drones, a big drone, to end-user applications. Notably, the small drones and big drone shown in Fig.~\ref{fig:drones} have been used in the experiments. The components and estimated prices of these drones are illustrated in Table~\ref{tab:drone_spec}. In the setup, four small drones and a big drone are formed in a cluster. Fig.~\ref{fig:drones} also shows an image captured by a drone (Autel robotics Evo II) using a 6K resolution camera at the height of 60m from the ground. This proves that a human detection algorithm is required as it is not easy to detect persons in the image by human eyes for a short moment in terms of seconds. When the drone flies higher, it is more challenging to detect the persons. 

\begin{figure}[t!]
	\centering
	\includegraphics[scale=0.22]{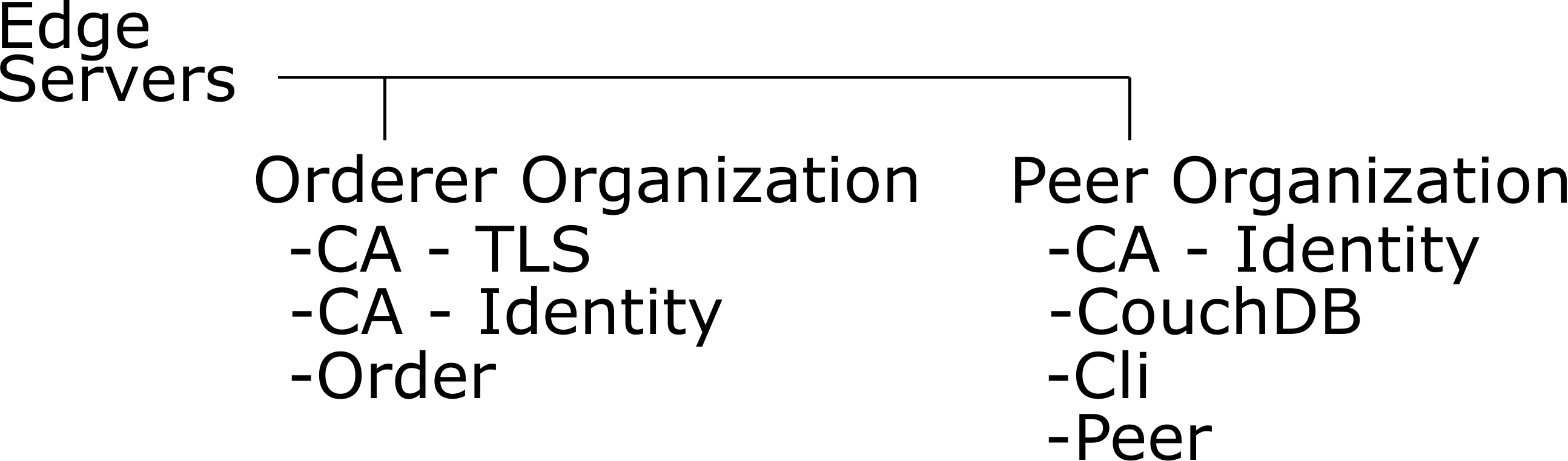}
	\caption{The organization structure of edge servers}
	\label{fig:orgStructure}
\end{figure}

\begin{figure}[t!]
	\centering
	\includegraphics[scale=0.08]{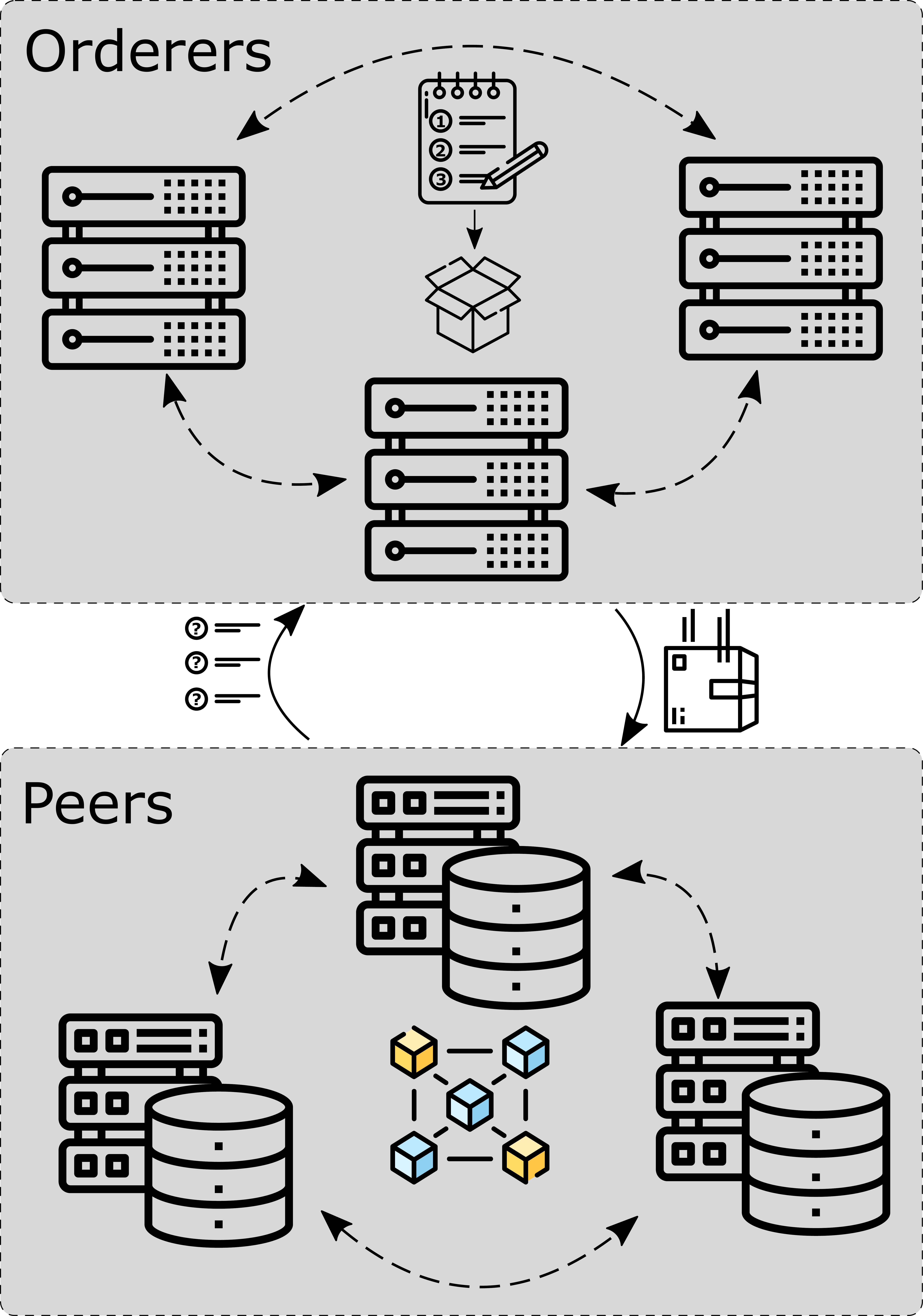}
    \caption{The design of edge servers' communication with the orderer organization and the peer organization}
    \label{fig:edgeCommunication}
\end{figure}

The network of edge servers encompasses six virtual machines, including three machines for Orderers tasks and three machines for Peers tasks. Also, this setup is based on a virtual machine tool called VirtualBox Version 5.2.22 r126460 (Qt5.6.2) on a physical machine with Intel(R) Core(TM) i7-8700 CPU @ 3.20GHz, 3.19GHz, 32GB RAM, and 64-bit OS, x64-based processor. 
Each edge server can desire to run Orderer or Peer tasks; even each edge server can run both services. However, to simplify tasks in a server, we desire to separate those services into different virtual machines with hardware components in Table~\ref{tab:harwareNetwork}. Each organization has specific tasks, as in Fig.~\ref{fig:orgStructure}, but notice that the certificate authority (CA) is unique for each organization. The detail for our edge servers' communication is specific in Fig.~\ref{fig:edgeCommunication}.

\begin{table}[]
    \centering
    \caption{The hardware components for edge servers}
    \label{tab:harwareNetwork}
    \begin{tabular}{p{0.85cm}|p{4.5cm}|c}
    \hline
                & Information                               & OS/Arch       \\ \hline    
    Docker      & Version: 18.09.7                          &  linux/amd64  \\ \hline
    Orderer     & Process: 2, 2GB Memory, fabric:2.2.1              &  linux/x86\_64 \\ \hline
    Peer        & Process: 4, 4GB Memory, fabric:2.2.1                 &  linux/x86\_64 \\ \hline
    \end{tabular}
 
\end{table}

\begin{figure*}[t!]
	\centering
	\includegraphics[width=0.92\textwidth]{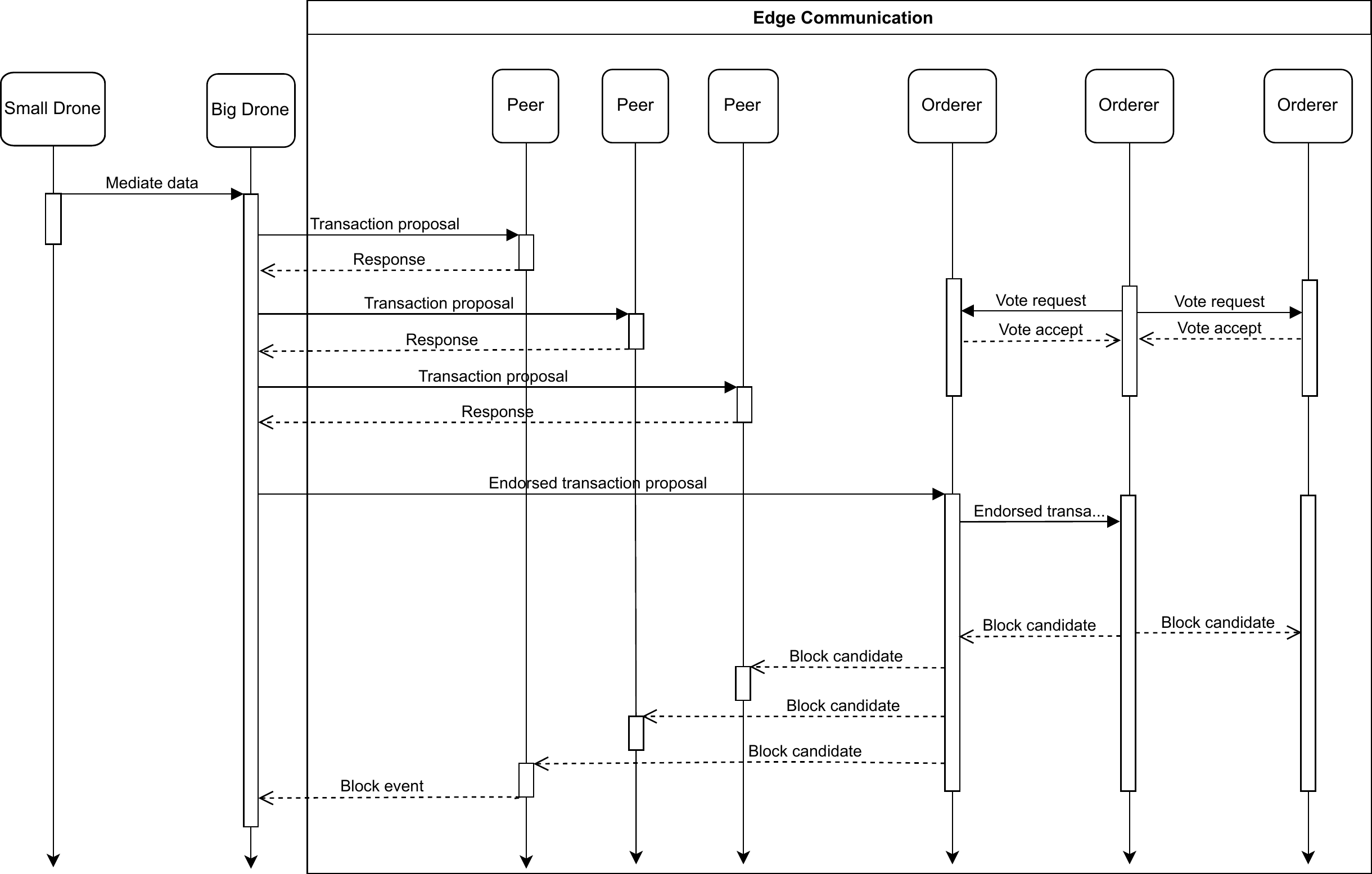}
	\caption{The sequence diagram of the proposal}
	\label{fig:sequenceDiagram}
\end{figure*}

\begin{table}[!h]
\caption{Drone specifications}
    \centering
    \begin{tabular}{p{1.4cm}|p{4.2cm}|p{1.4cm}}
    \hline
    Type & Main components & Price (Euros)  \\
    \hline
    Small drone  & & about 1400 \\
    &holybro x500 frame and properler & 100 \\
     & 4 Holybro ESC \& 4 880kV motors& 200 \\
     & Flight controller Pixhawk 4 & 220 \\
     & Lidar ranging sensor TFmini & 50 \\
     & Intel Realsense D435 camera  & 200 \\
     & GPS Pixhawk 4 + Wi-Fi Dlink + Teleradio & 100\\
     & Basgibg 5500mAh, 14.8V battery & 70\\
     &Intel UP square board & 250\\
     & Other components & 100 \\
     \hline
     Big drone && about 9.1k\\
     &DJI S1000 drone & 2k \\
     &DJI FLIR Zenmuse XT2 (Thermal + visual camera)& 6k\\
     & Nvidia Jetson Xavier NX & 400\\
     & Tattu 26000mAh Lipo Battery Pack & 600\\
     & Other components & 100\\
     \hline
    \end{tabular}
    \label{tab:drone_spec}
\end{table}

\begin{figure}[t!]
	\centering
	\includegraphics[scale=0.75]{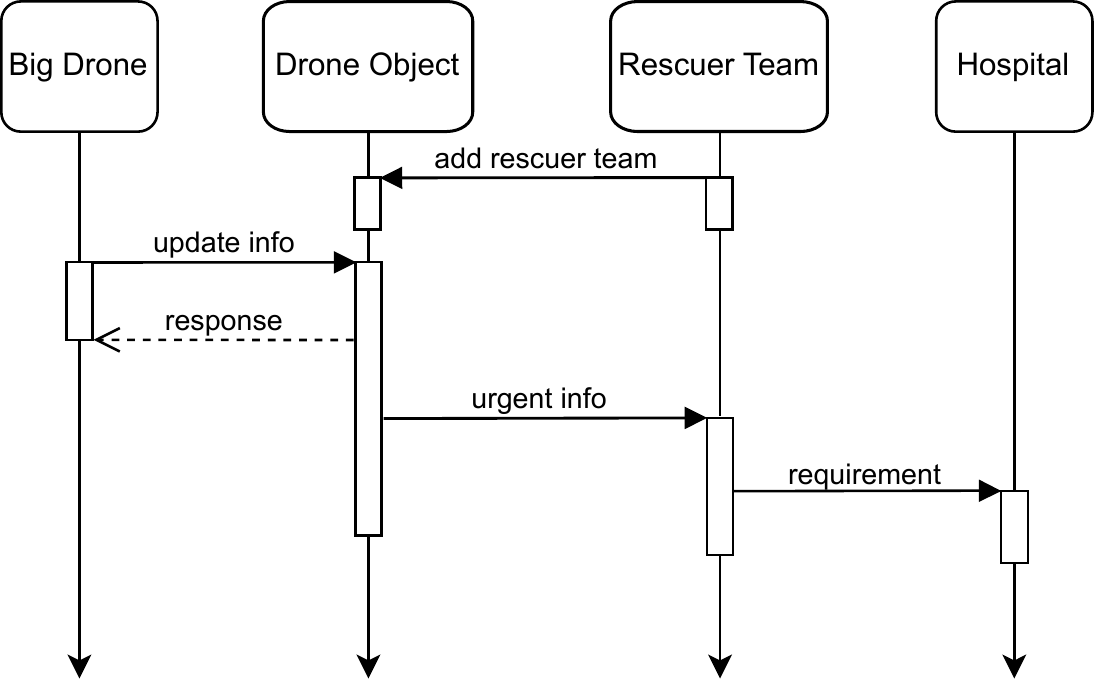}
	\caption{The sequence diagram of smart contract in the proposal}
	\label{fig:smartContract}
	\vspace{-10 pt}
\end{figure}

The blockchain layer is based on Hyperledger Fabric, a permissioned blockchain-based smart contract platform~\cite{androulaki2018hyperledger}. The system desires to adopt complete blockchain components, including a consensus mechanism, data model, smart contract flow, and cryptographic schemes by the Hyperledger Fabric. Since boat servers are the blockchain participants without limited resources, a fully functioning blockchain fits further developments, especially SAR systems that are doing tasks parallelly. Also, the permissioned blockchain provides a faster gaining agreement among participants and victim's confidentiality. In detail, Fabric introduces execute-order-validate architecture with a Raft consensus, key-value data model, and CA. Fabric asks the client to send transaction proposals to specific peers as known as endorsed peers, for executing and responding endorsed transaction proposals to the client in the execution phase. The client then submits an endorsed transaction proposal to an orderer who operates a total order and packs transactions into a block in the ordering phase. Finally, the orderer broadcasts the block candidate to all peers for validation at the validation phase. This procedure can be observed from Fig.~\ref{fig:sequenceDiagram} and the edge communication part of Fig.~\ref{fig:edgeCommunication}.
Hence, Fabric decomposes a blockchain-based system into sub-roles, including CA, ordering service nodes, peers, and clients. Due to a permissioned blockchain, Fabric utilizes the CA system to create the certificate and gain the authority for the system. The ordering service nodes as orderers implement the total order for transactions and ordered transactions through a Raft consensus. The peer stores blockchain, system state, and smart contract execution while the client task is about interaction with peers.
The edge servers deploy the blockchain system through contribution and verification based on Hyperledger Fabric in the proposed scenario. The edge servers process and form a CA, consensus, system state, and ledger storage. The big drones act as clients while other edge servers split and run peer and orderer roles as aforementioned. 
Due to the benefit of smart contracts, the design for the smart contract is a demand in the scenario. The smart contract sample consists of three prominent components, including drone object, rescue team, and hospital, as Fig.~\ref{fig:smartContract}. The drone object smart contract is to receive information from big drones that can be the location of drones, the connection between a big drone and small drones, and the history of previous data. The rescue team contract is a data structure of rescue team information, such as specialists, location, and previous records or archives. The hospital contract represents a list of hospital information like the rescue team's information and some extra information related to special equipment and suppliers. Therefore, once finding a victim in an emergency, a small drone sends information to a big drone who forwards information to the drone object. The system fast reacts and contacts the closest rescue team or the rescue team with available specialists by estimating current cases' priority. Besides, suppose the case is the most urgent and needs the support or preparation of a hospital. In that case, the system, through the smart contract, requests the closest hospital with support, such as emergency rooms, laboratories, and blood suppliers.

\section{Experimental results and analysis} \label{sec:experiment}
Latency of sending data (e.g., stream and images) by Wi-Fi from a small drone to a big drone is measured. In reach measurement, 100 transmitting times are carried out, and the mean of them is reported. In addition, different packet sizes such as 10k and 65508 bytes are sent in User Datagram Protocol (UDP). The latency results are shown in Table~\ref{tab:round_trip_latency}. As each transmitted video frame has a size of 2.3MB, it takes around 200-300ms to send a frame from a small drone to a big drone. The transmitted latency can vary significantly in terms of dozens of milliseconds as it is affected by different parameters, including noise source. It is difficult to measure the accuracy latency in actual applications, especially when drones are running. Therefore, a latency of 300ms is used in offloading decision-making.   

\begin{table}[!ht]
\caption{Latency for sending packets}
\label{tab:round_trip_latency}
\centering
\begin{tabular}{l|c|c}
\hline
Transfer protocol       & Round Trip Latency (ms)& Latency (ms)  \\ \hline
UDP (10k bytes)         & 7.4           & 3.7  \\ \hline
UDP (Max packet size)   & 11.3          & 5.6   \\ \hline
\end{tabular}

\end{table}

As mentioned in the experiments, state-of-the-art and also general image processing approaches for human detection such as Haar Cascades~\cite{viola2001}, HOG~\cite{dalal2005histogram}, YOLOv3-tiny~\cite{adarsh2020yolo}, YOLOv4~\cite{bochkovskiy2020yolov4}, and YOLOv4-tiny~\cite{jiang2020real} algorithms have been applied and evaluated with the proposed architecture. Using Haar Cascade classifiers was first introduced in~\cite{viola2001}. This method provides fast computation for object detection, but it is more sensitive to false-positive detection, 
HOG feature descriptors can be used with Support Vector Machine (SVM) or another machine learning algorithm as a classifier~\cite{dalal2005histogram}. HOG-based detector for detecting humans used in tests was from the OpenCV library. Frames are scaled down to 680x480 resolution for HOG and Haar Cascades approaches that are only applied in the smaller drone. For all versions of the YOLO algorithms, frames are resized to 416x416 resolution. Due to the scope of the paper, the mentioned approaches are tested with a video from the P-DESTRE dataset~\cite{kumar2020} to establish the feasibility of doing object detection tasks in smaller and bigger drones. Video is pre-processed to have similar technical specifications as the video output from an Intel RealSense D435 camera attached to the small drone. The technical specification is 1280x720 resolution, 30 fps framerate, 8073kB/s bitrate, and 30-second video length. 

Some results of running Haar Cascades, HOG, and YOLO3-tiny based algorithms at a small drone's Intel UP board are shown in Table~\ref{tab:fps_board_approach}. The result shows that the board requires large processing latencies for running these algorithms. 

In the case of YOLOv3-tiny, the latency of processing a frame at a small drone is around 710ms. In contrast, the total latency of transmitting a frame from a small drone to a big drone (i.e., transmission latency of 300ms) and processing a frame at a big drone (i.e., processing latency of 15-20ms) is around 320ms. Therefore, it is recommended to offload the computation tasks from a small drone to a big drone to minimize the total latency.

In the case of using Haar Cascades and HOG, the latency of processing a frame at a small drone is around 270ms and 330ms, respectively. The total latency of transmitting a frame from a small drone to a big drone and processing a frame on a big drone is around 310ms. In this situation, the system latency cannot be saved much when offloading. Even in the best case, when a frame transmission latency is around 200ms, the total latency cannot be reduced dramatically (e.g., around 60-100ms). However, offloading to a big drone provides the possibility to process the images with human detection algorithm with a high mean average precision (mAP) while maintaining low latency. 
Therefore, to simultaneously minimize the latency and achieve highly accurate human detection results, it is recommended to apply computation offloading from a small drone to a big drone according to the proposed architecture and the experimented drones.
The idle/standby power consumption of the Intel UP Squared board and Jetson Xavier NX is 3100mW and 4655mW, respectively. The energy consumption of an Intel UP Squared board is about 2083mJ, 2360mJ, and 5600mJ for processing a frame with Haar Cascades, HOG, and YOLOv3-tiny, respectively, while energy consumption for sending the frame is approximately 650-975mJ depending on the transmission latency (i.e., 200-300ms~per~frame). Therefore, offloading the processing from a small drone to a big drone helps reduce the energy consumption of the small drone significantly.

As mentioned, Jetson Xavier NX is used as a computation board for our big drone. Jetson Xavier NX supports several modes for enabling different power usage and the number of cores that can impact the offloading decision-making in terms of latency. The latency and power consumption results of running different human detection algorithms with these modes are shown in Table~\ref{tab:fps_board_approach} and Table~\ref{tab:energy_consumption_jetson}, respectively. The power consumption measurement results can contain an error rate of 1-5\% due to the error rate of the power monitoring device. The results show that the different latency for processing a frame in various power modes on the Jetson Xavier NX is around 5-7ms, while the different power consumption of these nodes is around 4-5W. Therefore, it is unnecessary to run the Jetson Xavier NX with high power modes and with the maximum number of CPU cores. 

The big drone without any computation board consumes about 1500W in a hover mode, and this can reach 4000W for the maximum power consumption. The Jetson Xavier NX consumes about 13-19W for running different human detection algorithms. The power consumption of the Jetboard is too small when comparing with the power consumption of the big drone. 
Therefore, applying computation offloading from small drones to the big drone almost does not affect the working time of the big drone which is around 20 minutes.     






\begin{table}[]
\caption{Intel UP Squared and Jetson Xavier NX performance for different approaches}
\label{tab:fps_board_approach}
\begin{tabular}{lccl}

\textbf{Intel UP Squared} & \multicolumn{1}{l}{} & \multicolumn{1}{l}{} &  \\ \hline
\multicolumn{1}{l|}{Approach} & \multicolumn{1}{c|}{\begin{tabular}[c]{@{}c@{}}Process Time (s)\end{tabular}} & \multicolumn{2}{c}{Frames per Second} \\ \hline
\multicolumn{1}{l|}{Haar Cascades} & \multicolumn{1}{c|}{239} & \multicolumn{2}{c}{3.7} \\ \hline
\multicolumn{1}{l|}{HOG} & \multicolumn{1}{c|}{271} & \multicolumn{2}{c}{3.3} \\ \hline
\multicolumn{1}{l|}{YOLOv3-tiny} & \multicolumn{1}{c|}{604} & \multicolumn{2}{c}{1.4} \\ \hline
 & \multicolumn{1}{l}{} & \multicolumn{1}{l}{} &  \\
\textbf{Jetson Xavier NX} & \multicolumn{1}{l}{} & \multicolumn{1}{l}{} &  \\ \hline
\multicolumn{1}{l|}{\multirow{3}{*}{Approach}} & \multicolumn{3}{c}{Frames per Second} \\ \cline{2-4} 
\multicolumn{1}{l|}{} & \multicolumn{1}{c|}{\begin{tabular}[c]{@{}c@{}}10W\\4 CPU Cores\end{tabular}} & \multicolumn{1}{c|}{\begin{tabular}[c]{@{}c@{}}15W\\4  CPU Cores\end{tabular}} & \multicolumn{1}{l}{\begin{tabular}[c]{@{}c@{}}15W\\6 CPU Cores\end{tabular}} \\ \hline
\multicolumn{1}{l|}{YOLOv3-tiny} & \multicolumn{1}{c|}{53.2} & \multicolumn{1}{c|}{61.8} & \multicolumn{1}{c}{67.8} \\ \hline
\multicolumn{1}{l|}{YOLOv4-tiny} & \multicolumn{1}{c|}{48.3} & \multicolumn{1}{c|}{57.5} & \multicolumn{1}{c}{61.3} \\ \hline
\multicolumn{1}{l|}{YOLOv4} & \multicolumn{1}{c|}{4.7} & \multicolumn{1}{c|}{4.7} & \multicolumn{1}{c}{4.7} \\ \hline
\end{tabular}
\end{table}


\begin{table}[]
    \centering
    \caption{Power consumption of an Intel UP Squared}
    \label{tab:energy_consumption_intel}
    \begin{tabular}{c|c}
    \hline
    Approach   &  Power Consumption (mW)\\ \hline
    Haar Cascades    &  7710\\ \hline
    HOG             & 7790\\\hline
    YOLOv3-tiny     &  7840\\
                      \hline    
    \end{tabular}
    
    \vspace{-10pt}
\end{table}

\begin{table}[]
    \centering
     \caption{Power consumption of Jetson Xavier NX when running the YOLOv3-tiny}
    \label{tab:energy_consumption_jetson}
    \begin{tabular}{c|c}
    \hline
    Power modes   &  Power Consumption (W)\\ \hline
 
   10W 2 CPU Cores   & 9.31 \\\hline
   10W 4 CPU Cores  & 13.11\\\hline
   15W 6 CPU Cores &  18.62\\ \hline
        
    \end{tabular}
   
\end{table}

\begin{table}[]
    \caption{The scalability of edge servers includes the average second being the time to handle a request and the capacity of messages}
    \label{tab:edge-scalability}
    \centering
    \begin{tabular}{l|c|c}
    \hline
    Request Types                       & Average second (s)   & Capacity (MB)          \\ \hline
    Invoke (update info)                &  0.672           &   ~4             \\ \hline
    Invoke (urgent info-requirement)    &  0.219           &   ~0.5           \\ \hline
    Query  (data drone)                 &  0.350           &   ~16            \\ \hline
    \end{tabular}
    \vspace{-10pt}
\end{table}

The experiment for edge server communication is to evaluate the throughput of the system in the previous section. In particular, the system deploys a set of smart contracts and calculates the throughput for two common requests, such as invoke and query. The invoke request is to make updates or changes to the system state, while the query request inquires information stored by the system. 
In this scenario, the invoke request attempts to update an image 1280x720 to the edge system corresponding to small drone images. The requests are executed by the peers before ordered and packed into a block candidate by orderers. The query request retrieves information at the current state; hence, the request does not need to communicate to the orderers. In the experiment, we evaluate the period for a request/query execution. In detail, the \textit{invoke (update info)} request represents the messages transferring images to the edge system. The \textit{invoke (urgent info-requirement)} is the big drone's request and communicates to all other smart contracts, as Fig.~\ref{fig:smartContract}, including a request from the big drone, urgent information message transfer, and requirement message transfer.
Since the invoke request needs several communications to peers and orderers, the period of invoke requests is higher than the query result, as in Table~\ref{tab:edge-scalability}. However, the invoke request related to smart contract communication takes a smaller period than the query request.
A possible reason for the higher period of processing a query request than an \textit{invoke (urgent info-requirement)} request is the time for retrieving information from the blockchain. Also, the query's capacity of approximately 16 MB is another reason. 
The capacity of update info invoke from a big drone to the system is 4MB for a mediate data update. A difference from those messages is the low capacity of function calls or requests among smart contracts. In addition, the evaluation is based on block configuration parameters, consisting of batch time out (2s), the maximum messages of a block (10 messages), and the maximum number of bytes (99MB). 





\section {Conclusion and Future work}   \label{sec:conclu}
This paper presented a novel IoD architecture utilizing blockchain for SAR missions. The architecture consisting of multi-layer drones (i.e., small drones and big drones) and edge computing supports computation offloading for prolonging the operating time of drones. In addition, the proposed architecture can help cover the large searching areas. Correspondingly, a rescue team can search for many people simultaneously in a large area with minimal time. Furthermore, the proposed architecture offers a high level of transparency, traceability, data integrity, and security via the integrating Hyperledger blockchain network. With the smart contract, many activities related to post SAR missions, such as operations, can be carried out quickly without interfering or controlling the control center. Via the experimented results, the proposed architecture is a promising approach to improve the quality of services of an Internet-of-Drones system.   

Although computation offloading can help prolong the operating time of small drones, the efficiency is not high. A drone uses most of the energy for running its motors. Therefore, it is required to combine computation offloading with techniques for reducing the energy consumption of the motors. In future work, the energy consumption of motors in different cases, including various heights, will be considered together with the energy consumption model for offloading. In addition, different zoom levels of a big drone's camera will be applied in the experiments to find out the approach that helps improve the quality of images without flying the big drone for a long distance. Also, the small drones and the big drones will be used for collecting images and videos of volunteers who perform several water-related activities such as swimming and floating on the water. The collected data will be sent to edge servers to train deep learning models. The trained model will be verified at big drones via detecting a person in the stages of SAR. 



\section{Acknowledgements}
Mr. Tri Nguyen is supported by TrustedMaaS project by the Infotech institute of the University of Oulu and the Academy of Finland 6Genesis Flagship (grant 318927).
%
%

\bibliographystyle{IEEEtran}
\bibliography{main}
\end{document}